\documentclass[%
 aip,
rsi,%
 amsmath,amssymb,
 reprint,%
]{revtex4-1}

\usepackage{graphicx}
\usepackage{dcolumn}
\usepackage{bm}

\usepackage{multirow}
\usepackage{hhline,booktabs,amsmath}
\usepackage{xcolor}

\begin{document}

\preprint{AIP/123-QED}

\title{Double magnetic phase transitions and magnetotransport anomalies in a new compound  Gd$_\textbf{2}$AgSi$_\textbf{3}$}

\author{Baidyanath Sahu}
 \altaffiliation{Corresponding author}
 
 \author{R. Djoumessi Fobasso}
  \homepage{Equal contributing author.}
 \affiliation{Highly Correlated Matter Research Group, Physics Department, University of Johannesburg, PO Box 524, Auckland Park 2006, South Africa}%
 
\author{Andr\'{e} M. Strydom}%
 
\affiliation{Highly Correlated Matter Research Group, Physics Department, University of Johannesburg, PO Box 524, Auckland Park 2006, South Africa}%

\begin{abstract}
Dc and ac$\textendash$magnetic susceptibility ($\chi$), specific heat ($C_\mathrm{P}$), electrical resistivity ($\rho$) and magnetoresistance measurements performed on the new polycrystalline compound  $\mathrm{Gd_2AgSi_3}$, crystallizing in the $\alpha$$\textendash$$\mathrm{ThSi_2}$ tetragonal structure, are reported. Two magnetic phase transitions were observed in dc and ac susceptibility, specific heat and resistivity measurements at temperatures $\mathrm{T_{N_1}} = 11$ K and $\rm{T_{N_2}} = 20$ K, despite a single site occupied by Gd atom, which is an indication of the complex magnetic behavior. $\mathrm{Gd_2AgSi_3}$ turns out be one of the rare Gd compound in which a minimum is observed in the temperature dependence of resistivity in the paramagnetic state and also negative magnetoresistance over a wide temperature range (above $\rm{T_{N_2}}$), mimicking the behavior of exotic  $\mathrm{Gd_2PdSi_3}$, in this ternary family. The isothermal magnetic entropy change, adiabatic temperature change, and refrigerant capacity reach a value of 9.5 J/kg-K, 7.5 K and 302 J/kg, respectively for the change of magnetic field 9 T.

 \end{abstract}


\maketitle

Rare-earth (RE) based ternary intermetallic compounds have been receiving considerable attention due to their interesting structural, magnetic and physical properties ~\cite{ST,R2TX3,HF,kondo,HF,kondo,MR,IV,ZTE,SG,MCE}.
In particular, heavy rare$\textendash$earth members, which were less considered for a long time in the literature due to the well$\textendash$localized nature of their 4f-electrons, yielded unique results in the $\rm{RE_{2}TX_{3}}$ (RE= rare-earth elements, T = transition metals, X = p-block elements) series of compounds. These compounds basically crystallize in the $\rm{ThSi_{2}}$ type of structure which has two modifications namely $\alpha$-$\rm{ThSi_{2}}$ and $\beta$-$\rm{ThSi_{2}}$ phases. Most of the $\alpha$-$\rm{ThSi_{2}}$-type of tetragonal structure belongs to a space group of ${I_{{4}_{1}}/amd}$~\cite{R2TX3}. However, compounds with $\beta$-$\rm{ThSi_{2}}$~mostly belong to space group $P6/mmm$ with $\mathrm{AlB_2}$-type of hexagonal structure. 
In contrast, Tran~\cite{Tran} has synthesized a series of $\rm{U_{2}TGa_{3}}$ (T = Ru, Rh, Ir, Pd, Au, Pt) compounds and found that they crystallize in the $\rm{CeCu_{2}}$  type of orthorhombic structure, which belongs to space group \textit{Imma}. Based on different structures, $\rm{RE_{2}TX_{3}}$ compounds show a variety of magnetic ground states, depending on indirect Ruderman-Kittel-Kasuya-Yosida (RKKY) exchange interaction between the ions.
This magnetic interaction dictates the physical properties of the 4f electrons giving rise to interesting and very often anomalous properties~ \cite{R2AgGe3, Skyr}.

$\rm{RE_{2}TX_{3}}$ series of compounds are subject to potential applications in magnetic refrigeration due to the exhibition of large magnetocaloric effect (MCE) \cite{Gd2NiSi3,Dy2PdSi3,Dy2CuSi3}. MCE$\textendash$based magnetic refrigeration techniques have attracted considerable attention due to their efficiencies and  environmentally friendly nature. The MCE is an intrinsic magneto$\textendash$thermodynamic phenomena and it describes the change of temperature of the magnetic material when subjected to variable magnetic fields. The MCE properties of a magnetic material are characterized in terms of isothermal magnetic entropy change ($\Delta S_m$) and adiabatic temperature change ($\Delta T_{ad}$). The refrigeration efficiency of magnetocaloric materials can be considered by evaluating refrigerant capacity (RC), and/or relative cooling power (RCP) \cite{book, Gd2ZnMnO6, rev1,rev2,rev3,deltaT1,deltaT2,RC}. The MCE of Gd$\textendash$itself and Gd$\textendash$based compounds are intensively studied to obtain a good magnetic refrigerant material for their large localised magnetic moments \cite{Gd2NiSi3,Gd2ZnMnO6,Gd,GdGeSi}. In this paper, we attempt to synthesize a new single$\textendash$phase intermetallic compound $\rm{Gd_{2}AgSi_{3}}$. Structural, magnetic, thermal, and electrical transport properties are systematically studied for the present compound. The  magnetocaloric effect  is extensively investigated from isothermal magnetizations and heat capacity.

\subsection{\label{sec:level2}Synthesis and Experimental Details}
A polycrystalline sample of $\mathrm{Gd_2AgSi_3}$ was synthesized by arc-melting the constituent elements of ultra high purity ($\geq$99.99 weight \%) under an inert Argon gas atmosphere on a water cooled copper hearth. The as-cast ingots were remelted several times to ensure homogeneity. The weight loss after the melting process was less than 0.5 weight \%. The as-cast sample was wrapped in tantalum foil, sealed in an evacuated quartz tube and annealed at 1323 K for one week to improve the homogeneity and followed by quenching into cold water. The phase purity of the annealed sample was checked by performing powder X-ray diffraction (XRD) experiments at room temperature using $\rm{Cu-K_{\alpha}}$ radiation of wavelength 1.54~\AA~ on a Rigaku powder diffractometer. The obtained XRD pattern was analyzed by Rietveld refinement method using the FULLPROF software \cite{rietveld, fullprof}. Temperature dependence of magnetization ($M(T)$) and field dependence of magnetization (M($H$)), were performed on a Physical Properties Measurement System (PPMS), Dynacool attached with a Vibrating Sample Magnetometer (VSM), Quantum Design, USA. $M(T)$ was recorded in the standard process of zero-field cooled (ZFC) and field cooled (FC) mode followed by measurements in field cooled warm-up mode (FW). In ZFC modes, the sample was first cooled to T = 2 K in zero field, then a field was applied and thereafter the data were recorded while warming. In FC modes, the sample was first cooled to 2 K in the presence of an applied field and the data were recorded during warming in the same field. The ac-susceptibility ($\chi_{ac}(T)$) measurement was carried out under different frequency with a constant driven field of 3 Oe using same Dynacool system. The specific heat $C_{\mathrm{P}}(T)$ measurement was carried out in the same Dynacool system using a thermal relaxation method in temperature range of 2$\textendash$300 K. The temperature dependent resistivity $\rho(T)$ and isothermal field dependence of resistivity $\rho(H)$ were measured by the standard four$\textendash$probe ac$\textendash$method using the same apparatus.

\subsection{\label{sec:level2}Results and Discussion}
\subsubsection{Structural properties}

\begin{figure}[h!]
      \centering
        \includegraphics[scale =0.32]{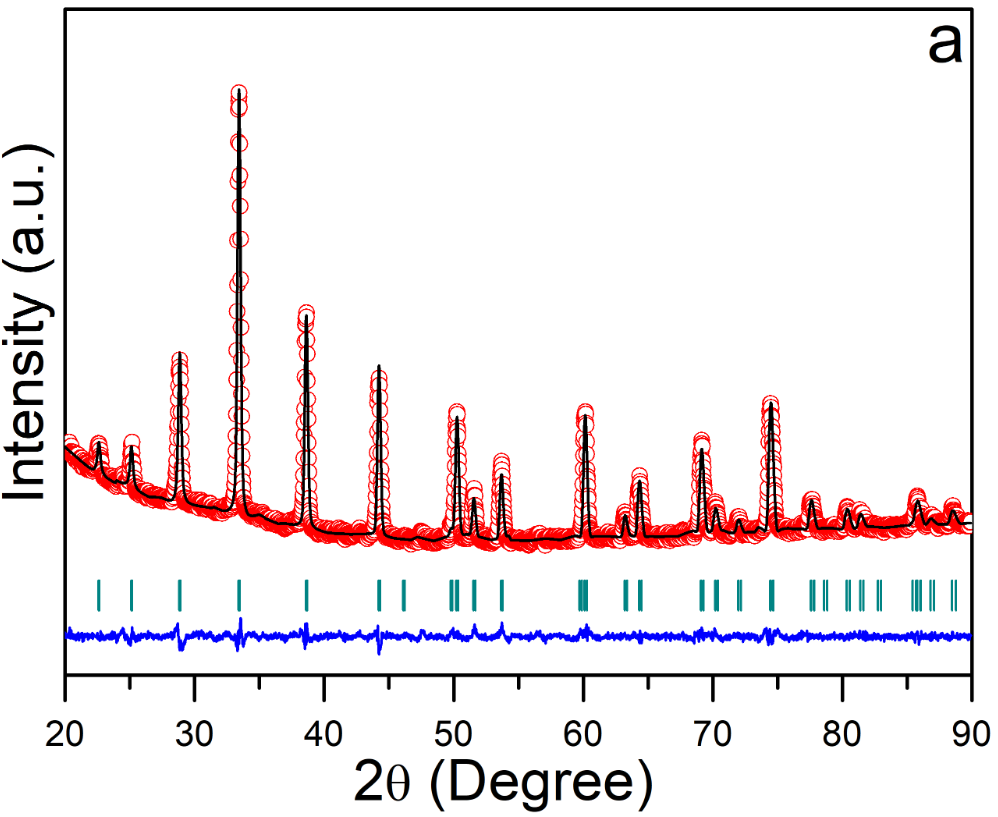}
       
      \vspace{1.0em}
        \begin{tabular}{cc}
        
        \includegraphics[scale =0.4]{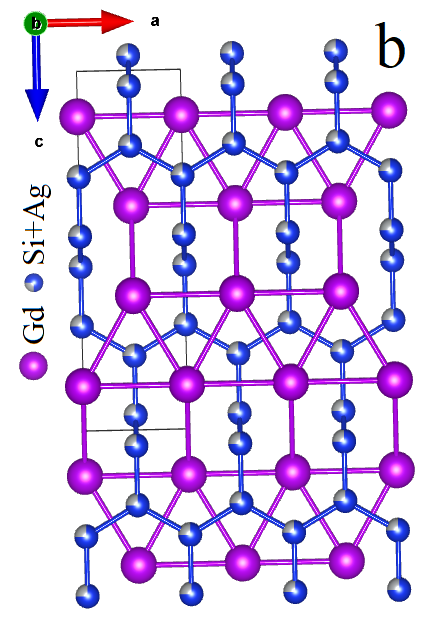}
         &
         \includegraphics[scale =0.4]{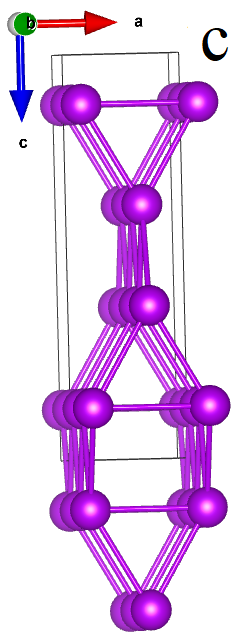}
         \end{tabular}
         \caption{(a): X-ray powder diffraction data for $\mathrm{Gd_{2}AgSi_{3}}$. Red symbols represent the experimental data and the black line represents the calculated data. The difference between experimental and calculated data is shown as blue line. A set of vertical bars represents the Bragg peak positions of the tetragonal $\alpha$-$\mathrm{ThSi_2}$ type structure. (b): The schematic representation of the tetragonal crystal structure of $\mathrm{Gd_2AgSi_3}$. (c): Gd atoms arrangement in the crystal structure.}
         \label{XRD}
\end{figure}

Fig.~\ref{XRD} shows the Rietveld refinement of the powder x-ray diffraction (XRD) data obtained for $\mathrm{Gd_2AgSi_3}$ performed using the centrosymmetric ${I_{{4}_{1}}/amd}$ space group. It is confirmed that the compound crystallizes in $\alpha$-$\mathrm{ThSi_{2}}$$\textendash$type of tetragonal structure. The schematic representation of this structure is shown in the inset of Fig.~\ref{XRD}. The crystallographic details obtained from the Rietveld refinement fit are given in table~\ref{refinedata}.

The smallest Gd$\textendash$Gd bond length in $\mathrm{Gd_2AgSi_3}$ is 4.0929(1) \AA. This value is larger than the expected Gd$\textendash$Gd bond (atomic radii)  of 3.5730 \AA~and therefore suggests a rather weak interaction between the rare-earth atoms. The shortest Gd$\textendash$M (M = Ag + Si) bond was found to be 3.1226(1) \AA. Similar values of RE-M bond (3.2508(2) \AA~for $\mathrm{Ce_2AgGe_3}$, 3.2311 \AA~for $\mathrm{Pr_2AgGe_3}$ and 3.1996 \AA~for $\mathrm{Nd_2AgGe_3}$)\cite{R2AgGe3} have been reported for other 2$\textendash$1$\textendash$3 compounds forming in the same $\alpha$-$\mathrm{ThSi_{2}}$ crystal structure with same space group. Each Gd atom is surrounded by 10 M nearest neighbours atoms.

\begin{table}
\caption{\label{comparisrisionMCE} The lattice parameters and unit cell volume $\mathrm{Gd_2AgSi_3}$ compound obtained from the Rietveld refinement of XRD patterns for tetragonal phase along with the atomic coordinate positions.}
\begin{ruledtabular}
\begin{tabular}{ccccc}
\hspace{-1.5 in} a 	& \hspace{-0.0 in}  & \hspace{-0.5 in} & \hspace{-0.1 in}  & \hspace{-0.7 in}4.094(2) \AA    \\
\hspace{-1.5 in} b 	& \hspace{-0.0 in}  & \hspace{-0.5 in} & \hspace{-0.1 in}  & \hspace{-0.7 in} 4.094(2) \AA \\
\hspace{-1.5 in} c  & \hspace{-0.0 in}  & \hspace{-0.5 in} & \hspace{-0.1 in}  & \hspace{-0.7 in} 14.178(2) \AA \\
\hspace{-1.5 in} V  & \hspace{-0.0 in}  & \hspace{-0.5 in} & \hspace{-0.1 in}  & \hspace{-0.7 in} 183.461(3)  \AA$^3$ \\
\hspace{-1.5 in} $\mathrm{R_\textit{{p}}}$ (\%) & \hspace{-0.0 in}  & \hspace{-0.5 in} & \hspace{-0.1 in}  & \hspace{-0.7 in} 17.7 \\
\hspace{-1.5 in} $\mathrm{R_\textit{{wp}}}$ (\%) & \hspace{-0.0 in}  & \hspace{-0.5 in} & \hspace{-0.1 in}  & \hspace{-0.7 in} 11.2 \\
\hspace{-1.5 in} $\mathrm{R_\textit{{exp}}}$ (\%) & \hspace{-0.0 in}  & \hspace{-0.5 in} & \hspace{-0.1 in}  & \hspace{-0.7 in} 5.69 \\
\hspace{-1.5 in} $\chi ^2$  & \hspace{-0.0 in}  & \hspace{-0.5 in} & \hspace{-0.1 in}  & \hspace{-0.7 in} 4.74 \\
\hline
\\
\hspace{0.1 in} Atomic coordinates  for $\mathrm{Gd_2AgSi_3}$ 	& \hspace{-0.2 in}  & \hspace{-0.2 in} & \hspace{-0.2 in}     & \hspace{-0.2 in} \\
\hline
\\
 \hspace{-1.5 in}Atom & \hspace{-2.5 in}Wyckoff & \hspace{-1.5 in}$x$ & \hspace{-0.7 in}$y$ & \hspace{-0.2 in}$z$ \\
 \hline
 & & & & \\
 \hspace{-1.5 in}Gd & \hspace{-2.5 in}$4a$ & \hspace{-1.5 in}0 & \hspace{-0.7 in}0.7500 & \hspace{-0.2 in}0.1250\\
 \hspace{-1.5 in}Ag & \hspace{-2.5 in}$8e$ & \hspace{-1.5 in}0 & \hspace{-0.7 in}0.2500 & \hspace{-0.2 in}0.2925(1)\\
 \hspace{-1.5 in}Si & \hspace{-2.5 in}$8e$ & \hspace{-1.5 in}0 & \hspace{-0.7 in}0.2500 & \hspace{-0.2 in}0.2925(1)\\
\end{tabular}
\end{ruledtabular}
\label{refinedata}
\end{table}

\subsubsection{Magnetic properties}
\begin{figure}[h!]
	\centering
	\includegraphics[scale = 0.30]{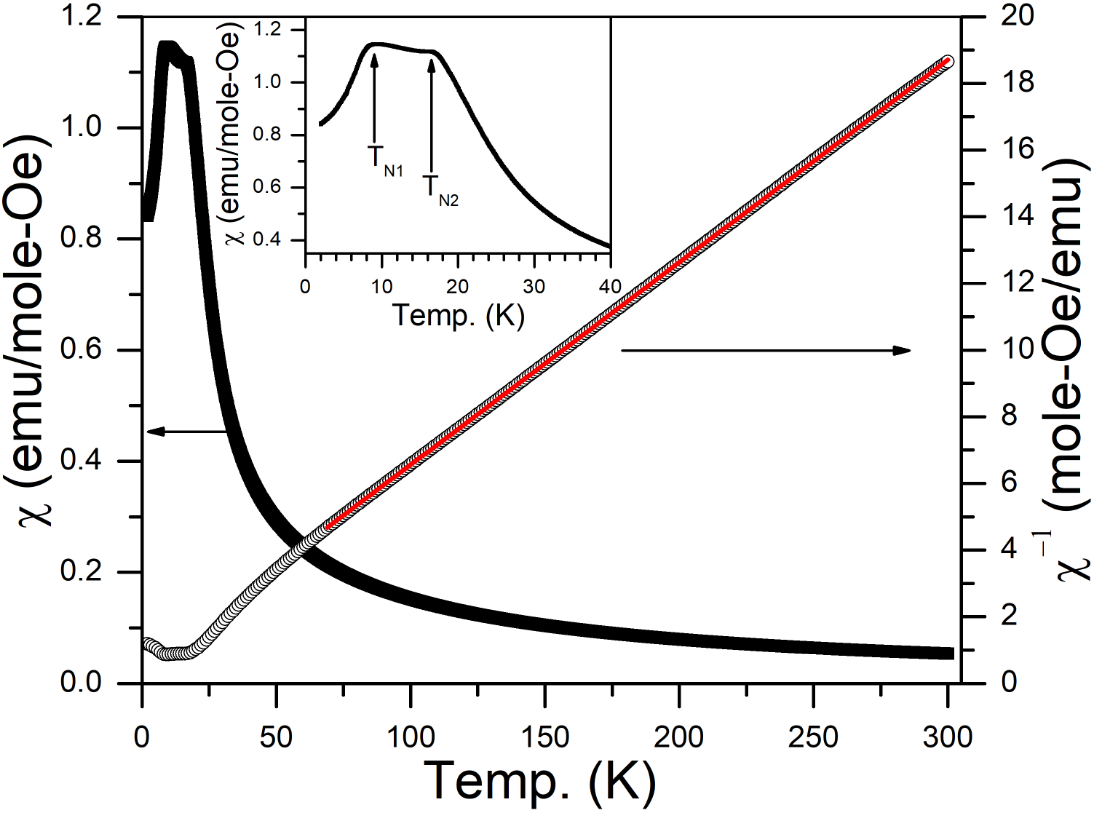}
	\caption{Left scale: temperature dependence of dc-magnetic susceptibility ($\chi_{dc} = M/H)$ of $\rm{Gd_{2}AgSi_{3}}$ measured under a magnetic field of 1.0 T in FC mode. Right scale: inverse magnetic susceptibility as function of temperature for the same data. Inset: Expanded region at low temperatures of $\chi_{dc}(T)$ data showing two anomalies at $\rm{T_{N_1}} = 11$ K and $\rm{T_{N_2}} = 20 $ K (see arrows).}
	\label{chivsT}
\end{figure}

Fig.~\ref{chivsT} shows the temperature variation of field-cooled dc-magnetic susceptibility ($\chi_{dc}(T)$ = $M(T)$/$H$) of $\mathrm{Gd_{2}AgSi_{3}}$ measured under an applied field of H = 1.0~T. Below 30 K, $\chi_{dc}(T)$ exhibits two anomalies. The expanded low - temperature region of $\chi_{dc}(T)$ is shown in the inset of Fig.~\ref{chivsT} to highlight the double magnetic transition in $\mathrm{Gd_{2}AgSi_{3}}$ . This double transition will be further investigated below.

The inverse magnetic susceptibility ($\chi^{-1}_{dc}(T)$) of the compound is presented on the right hand scale of Fig.~\ref{chivsT}. 
The $\chi^{-1}_{dc}(T)$ data below 300 K and 70 K can be described using the Curie$\textendash$Weiss law, $\chi_{dc}$ = C/($T - \mathrm{\theta_P}$), where C is the Curie constant proportional to the square of the effective magnetic moment ($\mu_{\mathrm{eff}}$) and $\rm{\theta_P}$ is the paramagnetic Weiss temperature. The least$\textendash$squares fit of the equation to the data leads to the values of $\mathrm{\mu_{eff}}$ = 8.12 $\mu_\mathrm{B}$/Gd and $\rm{\theta_P}$ = -8~K. The obtained $\mu_{\mathrm{eff}}$ value is comparable to the theoretical free-ion value for $\rm{Gd^{3+}}$ ion which is $\approx$ 7.94 $\mu_\mathrm{B}$/Gd. The negative value of $\rm{\theta_P}$ indicates that the dominant interactions in the compound are antiferromagnetic.

\begin{figure}[ht]
	\centering
	\includegraphics[scale = 0.35]{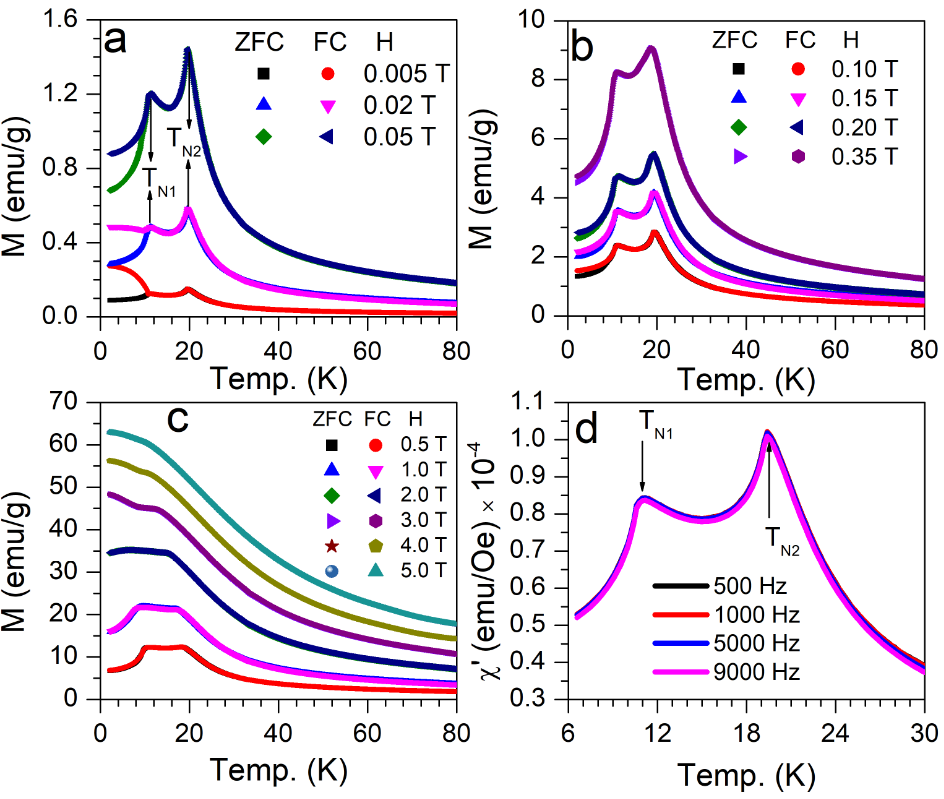}
	\caption{(a)$\textendash$(c):Temperature dependence of dc-magnetization measured under ZFC and FC protocols at different applied magnetic fields. (d): Temperature variation of the real part of ac-susceptibility measured at different ac-field frequencies with 3.0 Oe ac-driven field.}
	\label{dcandaclowT}
\end{figure}

Two anomalies are found at low temperature, one at 20 K and another at 11 K. In order to get an insight into the two magnetic transitions, temperature dependent magnetization of $\rm{Gd_{2}AgSi_{3}}$ in ZFC and FC modes were measured at different applied magnetic fields and are represented in Fig.~\ref{dcandaclowT}a~$\textendash$~\ref{dcandaclowT}c. The two transitions show antiferromagnetic behavior with N\'{e}el temperatures $\rm{T_{N_1}} = 11$ K and $\rm{T_{N_2}} = 20 $ K clearly visible in Fig.~\ref{dcandaclowT}(a). One can also see from Fig.~\ref{dcandaclowT}a~, and \ref{dcandaclowT}b that the $M(T)$ shows an irreversibility behaviour between ZFC and FC magnetization. However, the relative magnitude of the bifurcation gradually decreases with increase in applied magnetic field and the two branches overlap for field values above 0.5 T.

Fig.~\ref{dcandaclowT}(d) shows the real part of the $\chi_{ac}(T)$ data recorded at three different ac-field frequencies. The real component $\chi{'}_{ac}(T)$ of ac-susceptibility data confirm that the sample exhibits two magnetic transitions. However, there is no signature of spin glass behavior in the compound as the peaks in $\chi{'}_{ac}(T)$ data are frequency independent in the range of 500 to 9000 Hz.

\subsubsection{Specific heat}

\begin{figure}[ht]
	\centering
	\includegraphics[scale = 0.375]{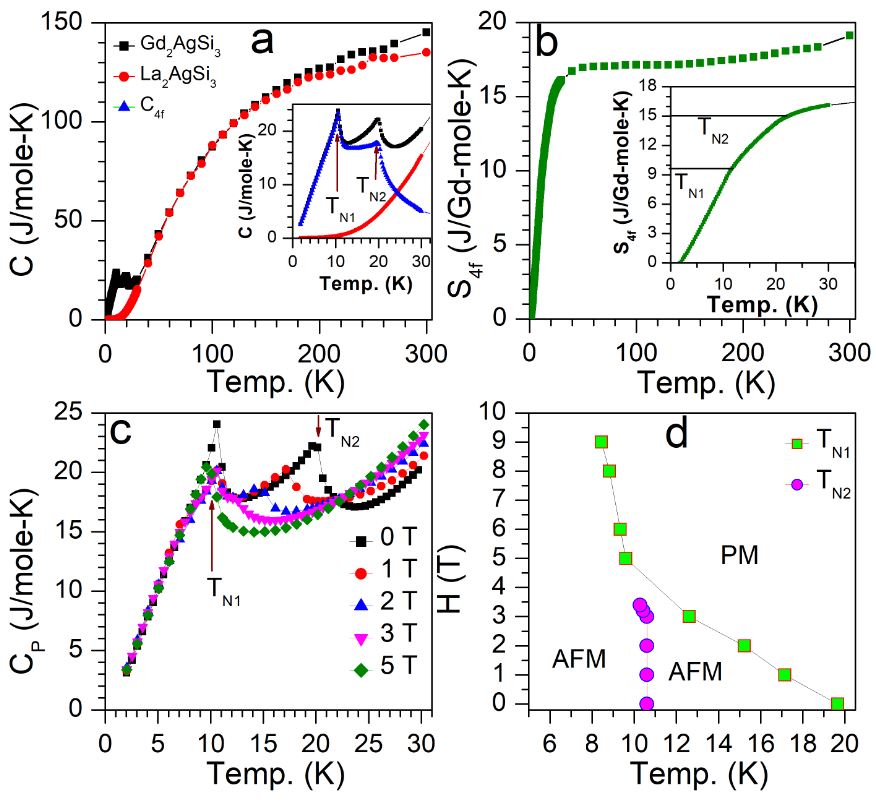}
	\caption{(a): Zero-field specific heat ($C_\mathrm{p}$) of $\rm{Gd_{2}AgSi_{3}}$ in black symbols. The temperature dependence of zero-field specific heat of $\rm{La_{2}AgSi_{3}}$ is represented as a red line. The magnetic contribution ($C_\mathrm{4f}$) of specific heat is in blue symbols. Inset represents the expanded low-temperature region of $C_\mathrm{p}$ of $\rm{Gd_{2}AgSi_{3}}$, $\rm{La_{2}AgSi_{3}}$ and $\rm{C_{4f}}$ of $\rm{Gd_{2}AgSi_{3}}$. (b): Temperature dependent magnetic entropy ($\rm{S_{4f}}$). Inset represents the expanded low temperature region of $\rm{S_{4f}}$. (c): The temperature dependence of $C_\mathrm{{p}}$ at different applied magnetic fields for $\rm{Gd_{2}AgSi_{3}}$. (d): Phase diagram for the variation of $\rm{T_{N_1}}$ and $\rm{T_{N_2}}$ with applied magnetic field.}
	\label{fig4}
\end{figure}

The temperature variation of specific heat ($C_\mathrm{P}(T)$) of $\rm{Gd_{2}AgSi_{3}}$ is presented in Fig.~\ref{fig4}(a) along with its isostructural non-magnetic analogue $\rm{La_{2}AgSi_{3}}$ (solid red line). Both compounds have $C_P$ values close to the Dulong-Petit value of 3nR $\approx$ 147 J/mole-K, where n is the number of atoms in the formula unit and R is the universal gas constant. The compound $\rm{La_{2}AgSi_{3}}$ exhibits a typical behavior for a non-magnetic metal between 2 K and 300 K. 

The inset panel in Fig.~\ref{fig4}(a) shows the expanded temperature region between 2 K and 30 K to highlight the double phase transition in $\rm{Gd_{2}AgSi_{3}}$. One can note that the low-temperature region of specific heat gives an evidence of two magnetic phase transitions with two nearby peaks at $\rm{T_{N_1}}$ $\approx$ 11 K and $\rm{T_{N_2}}$ $\approx$ 20 K, which is consistent with the peaks observed in $\chi(T)$ data. The 4f$\textendash$magnetic contribution of specific heat (${C_{4f}}$) was estimated by subtracting the zero-field specific heat for isostructural $\rm{La_{2}AgSi_{3}}$. The variation of $C_\mathrm{4f}$ as a function of temperature is shown in Fig.~\ref{fig4}(a).

The magnetic entropy $\rm{S_\mathrm{4f}}$ has been estimated using the term $\int(C_\mathrm{4f}/T)dT$. The variation of ${S_\mathrm{4f}}$ as function of temperature is shown in Fig.~\ref{fig4}(b) and the expanded region at low temperature is shown in inset of Fig.~\ref{fig4}(b). The magnetic entropy at $\rm{T_{N_1}}$ is ${S_\mathrm{4f}}$ $\approx$ 9.6 J/mole-Gd-K, and is ${S_\mathrm{4f}}$ $\approx$ 15 J/mole-Gd-K at $\rm{T_{N_2}}$. Here, ${S_\mathrm{4f}}$ shows only 2/3 of Rln(2S+1) is released at $\rm{T_{N_1}}$, while the entropy at $\rm{T_{N_2}}$ is sightly less than that Rln8, which is expected for the full multiplet of $\rm{Gd^{3+}}$.

The temperature dependence of $C_\mathrm{{p}}$ of $\rm{Gd_{2}AgSi_{3}}$ was measured under different fields up to 5 T in the temperature range of 30 K to 2 K. Fig.~\ref{fig4}(c) shows the $C_\mathrm{{p}}$  $vs.$ $T$ plots for different values of magnetic field. One can note from Fig.~\ref{fig4}(c) that the peak at $\rm{T_{N_2}}$  shifts to lower temperatures with increase in applied magnetic fields. 
This behavior is expected for an antiferromagnetic transition.
However, the peak position at $\rm{T_{N_1}}$ is not changing with fields up to 3 T. At a field value of 5 T, only one peak remains below $\rm{T_{N_1}}$ which might shifts marginally lower at this field and this transition evidently requires higher magnetic fields to be suppressed. The field \textit{vs.} temperature phase diagram of $\rm{Gd_{2}AgSi_{3}}$ derived from the field dependent specific heat measurement is shown in Fig.~\ref{fig4}(d). The specific heat result indicates that the compound has a complex magnetic structure with two phase transitions.

\subsubsection{Electrical resistivity}

\begin{figure}[h!]
	\centering
	\includegraphics[scale = 0.62]{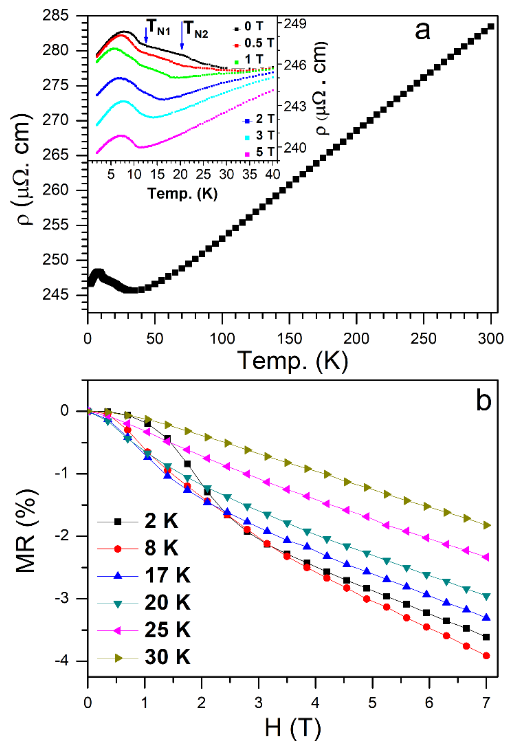}
	\caption{(a): Temperature dependence of resistivity ($\rho(T)$) for $\rm{Gd_{2}AgSi_{3}}$ in zero applied magnetic field. Inset: $\rho(T)$ was measured under different magnetic field in low temperature region. (b) Field dependence of the magnetoresistivity isotherms of
	$\rm{Gd_{2}AgSi_{3}}$.}
	\label{fig5}
\end{figure}

The zero-field electrical resistivity $\rho (T)$ of  $\rm{Gd_{2}AgSi_{3}}$ is shown in Fig.~\ref{fig5}(a). The low-temperature region is expanded and shown in the inset of Fig.~\ref{fig5}(a). The $\rho (T)$ gradually increases with temperature above 40~K. This indicates that the compound shows ordinary metallic behaviour in the paramagnetic region. However, $\rho (T)$ shows a broad minimum at T~$\approx$ 35~K and a maximum at T~$\approx$ 8~K before it decreases. The anomalies at below the transition temperature is possible for the superzone effects \cite{superzone, superzone1} 

Fig.~\ref{fig5}a shows two kinks in $\rho (T)$ at about 21 K and 10 K (see arrows), marking the onset of magnetic ordering. These transition temperatures are consistent with the $\chi (T)$ and $C_{p}$ results. It is also observed that the $\rho (T)$ does not drop immediately below either $\rm{T_{N_1}}$ or $\rm{T_{N_2}}$, which would be expected due to the loss of spin-disorder scattering \cite{NegativeMR}. This behavior highlights the complexity of the magnetic structure in this compound which also reflects from the formation of superzone boundary gaps in some portions of the Fermi surface~ \cite{superzone}. The inset of Fig.~\ref{fig5}(a) shows the $\rho (T)$ of $\rm{Gd_{2}AgSi_{3}}$ for different values of applied magnetic fields. 
As seen in the inset panel of Fig.~\ref{fig5}a, both the transition temperatures $\rm{T_{N_1}}$ and $\rm{T_{N_2}}$ are getting suppressed with increase in magnetic field, which is commonly seen in antiferromagnetic materials. However, the superzone feature remains visible upto 5 T. It can also be noted that $\rho (T)$ decreases with increasing magnetic field which indicates a negative magnetoresistance (MR) feature\cite{NegativeMR}.

In order to obtain the magnetic field dependent magnetoresistance (MR), isothermal field dependent resistivity measurements were carried out at different temperature in the range of 2~K to 30 K. MR was estimated by using the following formula:
\begin{align}
\rm{MR} = \frac{\rho(H,T)-\rho(0,T)}{\rho(0,T)}\times 100\%,
\end{align}
where $\rho(0,T)$ is the resistivity at zero magnetic field and
$\rho(H,T)$ is the resistivity at applied field 'H'. Fig.~\ref{fig5}(b) shows the nature of field dependent MR. It is apparent from Fig.~\ref{fig5}(b) that $\rm{Gd_{2}AgSi_{3}}$ exhibits negative MR over a wide range of temperature. Negative MR is also reported on other rare-earth based antiferromagnetic compound~\cite{superzone, NegativeMR, NegativeMR1, NegativeMR2, NegativeMR3}. 

\subsubsection{Magnetocaloric effect}

\begin{figure}[ht]
	\centering
	\includegraphics[scale = 0.34]{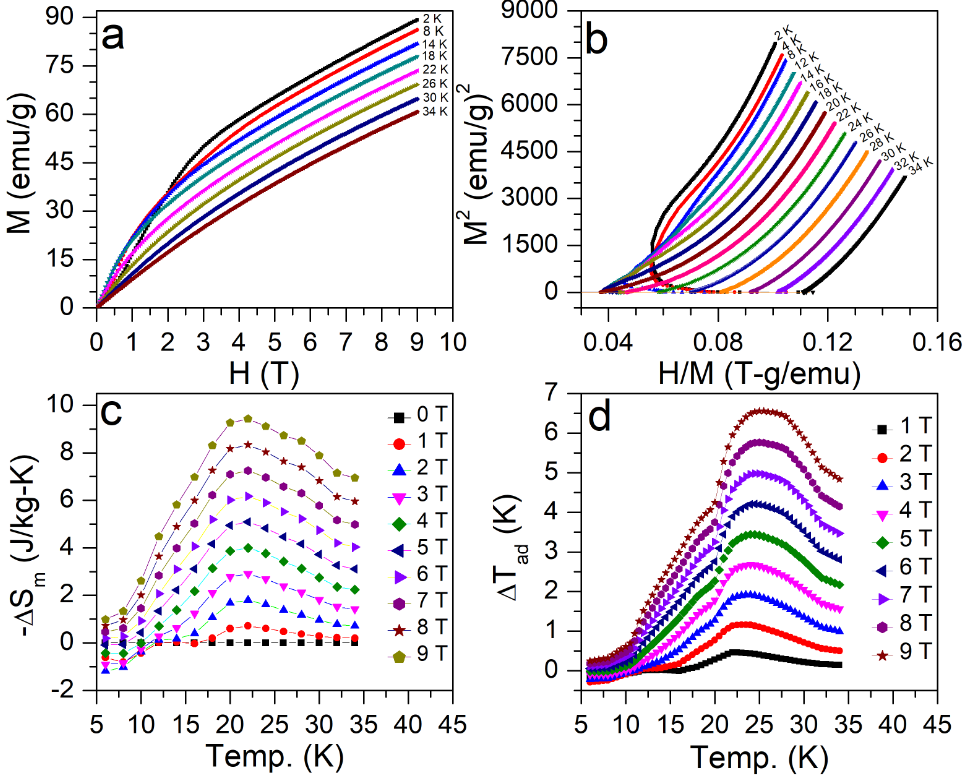}
	\caption{(a): Isothermal magnetization during magnetic field change 0 T $\rightarrow$ 9 T of $\rm{Gd_{2}AgSi_{3}}$ at different temperatures between 2 K and 34 K with a step of 2 K. (b) Arrott plots ($\rm{M^{2}}$) \textit{vs.} H/M at different temperatures. (c) Temperature variation of magnetic entropy change ($\Delta S_m$) for different fields. (d) Variation of adiabatic temperature change as a function of temperature for different fields.}  
	\label{Fig6}
\end{figure}

In order to obtain the isothermal magnetic entropy changes for MCE study and for more confirmation about order of magnetic phase transition, isothermal magnetization for different temperatures has been measured for $\rm{Gd_{2}AgSi_{3}}$. There was no hysteresis loop at 2 K (not shown), which confirmed that the compound shows soft magnetic behavior. Fig.~\ref{Fig6}(a) shows magnetization as function of field $M(H)$ at different temperatures between 2 K and 34 K with a step of 2 K for field cycle 0 T $\rightarrow$ 9 T. The spontaneous magnetic moment of $\rm{Gd_{2}AgSi_{3}}$ at 2 K for a field of 9 T is 5.6 $\mu_B$/$\rm{Gd^{3+}}$, which is less than the spontaneous magnetization of free $\rm{Gd^{3+}}$ (gJ = 7$\mu_B$) moments. 
One can see that isothermal M$\textendash$H shows a typical metamagnetic behaviour below $\rm{T_{N_1}}$. The critical magnetic field was determined from the maximum of d$M$/d$H$ and found to be 0.9 T for 2 K.
The Arrott$\textendash$plots ($M^2$ $\textit{vs.}$ $H/M$) was derived from the isothermal $M(H)$ curves to investigate the nature of the phase transitions. Fig.~\ref{Fig6}(b) depicts the Arrott plot for $\rm{Gd_{2}AgSi_{3}}$ compound. 
Based on the Banarjee criterion~\cite{banarjee}, the $M^2$ \textit{vs.} $H/M$ plots show negative slopes as a function of magnetic field, indicating a first-order magnetic phase transition. 

\begin{table}[ht]
\caption{The transition temperature, the maximum values of
magnetic entropy change ($\rm \Delta \textit{S}_{M}$), and refrigeration capacity (RC) under the field change of $0\textendash5~\rm T$ for some rare-earth compounds of $\mathrm{RE_2T_2X}$.}
\centering 
\begin{tabular}{ccccc}
\hline\hline
\\  
Method&  \hspace{0.2cm}$\rm{\textit{T}_{N}}$/$\rm{\textit{T}_{C}}$& \hspace{0.7cm}$\rm -\Delta \textit{S}_{M}$& \hspace{0.2cm}RC& \hspace{0.3cm}Ref  \\ [0.5ex]
       & (K) &\hspace{0.5cm}(J/kg.K) & \hspace{0.3cm}(J/kg)&     \\ [0.5ex]
\hline
\\ 

\hspace{0.0cm}$\rm{Gd_{2}NiSi_{3}}$& \hspace{0.3cm}16.4& \hspace{0.3cm}$\approx$13& \hspace{0.3cm}$\approx$300$^*$& \hspace{0.6cm}\cite{MCE} \\

\hspace{0.0cm}$\rm{Gd_2ZnMnO_6}$& \hspace{0.3cm}6.4& \hspace{0.3cm}15.17&  \hspace{0.3cm}226.2$^*$&  \hspace{0.6cm}\cite{Gd2ZnMnO6}  \\

\hspace{0.0cm}$\rm{TmZn}$& \hspace{0.3cm}8.4& \hspace{0.3cm}26.9&  \hspace{0.3cm}214&  \hspace{0.6cm}\cite{TmZn} \\

\hspace{0.0cm}$\rm{HoPdIn}$& \hspace{0.3cm}23/6& \hspace{0.3cm}14.6&  \hspace{0.3cm}374&  \hspace{0.6cm}\cite{HoPdIn} \\

\hspace{0.0cm}$\rm{Ho_3Rh_2}$& \hspace{0.3cm}23& \hspace{0.3cm}14.1&  \hspace{0.3cm}382&  \hspace{0.6cm}\cite{Ho3Rh2} \\

\hspace{0.0cm}$\rm{GdCo_2B_2C}$& \hspace{0.3cm}17.2& \hspace{0.3cm}10.3&  \hspace{0.4cm}238$^*$& \hspace{0.6cm}\cite{GdCo2B2C} \\

\hspace{0.0cm}$\rm{Gd_2Rh_3Ge}$& \hspace{0.3cm}64& \hspace{0.3cm}9.4&  \hspace{0.4cm}352$^*$& \hspace{0.6cm}\cite{Gd2Rh3Ge} \\

\hspace{0.0cm}$\rm{Ho_2FeAlO_6}$& \hspace{0.3cm}$\textless$2& \hspace{0.3cm}7.1&  \hspace{0.3cm}144.5&  \hspace{0.6cm}\cite{Ho2FeAlO6} \\

\hspace{0.0cm}$\rm{Pr_2Ni_{0.95}Si_{2.95}}$& \hspace{0.3cm}3.3& \hspace{0.3cm}$\approx$6&  \hspace{0.3cm}---&  \hspace{0.6cm}\cite{Pr2NiSi3} \\

\hspace{0.0cm}$\rm{Gd_{2}AgSi_{3}}$& \hspace{0.3cm}20& \hspace{0.3cm}5.1& \hspace{0.3cm}103&  \hspace{0.6cm}This work  \\ [1ex]
\hline 
$^*$ represents for RCP.
\end{tabular}
\label{compare} 
\end{table}

Double phase magnetic materials are interesting for MCE as they show a wide range of maximum magnetic entropy change ($\Delta S^{max}_m$) \cite{dualphase,twopeaks}. The broad behaviour of $\Delta S^{max}_m$ are beneficial for the application of magnetic refrigeration \cite{dualphase,twopeaks,Eu4PdMg}. Hence, the MCE of $\mathrm{Gd_2AgSi_3}$ has been investigated and the values of $\Delta S_m$ have been calculated from magnetization isotherms using the Maxwell's thermodynamic relation~\cite{book}:
\begin{eqnarray}
\Delta S_{m} (T,H) = \int\limits_{0}^{H} \left(\frac{\partial M}{\partial T}\right)dH.
\end{eqnarray}
Fig.~\ref{Fig6}(c) shows the temperature dependence of $- \Delta S_m$ for the change of magnetic field up to 9 T. The negative values of $- \Delta S_m$ below $\rm{T_{N_2}}$ for small magnetic field change is possible for the antiferromagnetic ground state of $\mathrm{Gd_{2}TX_{3}}$. Positive value of $- \Delta S_m$ at the relatively large magnetic field (H $>$ 5 T) was observed for the possibility of field$\textendash$induced antiferromagnetic$\textendash$ferromagnetic metamagnetic transition~\cite{DyNiGa}. The $- \Delta S_m$ $vs.$ T graph shows a peak at around the magnetic transition temperature $\rm{T_{N_2}}$. Here, we have observed a single  peak instead of two peaks, which might be possible due to the presence of a metamagnetic behavior upto first transition. A maximum value of $- \Delta S^{max}_m = $ 9.5 J/kg$\textendash$K is found around the transition temperature $\rm{T_{N_2}}$ at a change of magnetic field of 9 T. The observed value is compared with other reported values in Table~\ref{compare}.

$\Delta T_{ad}$ was evaluated from both $- \Delta S_m$($T$, $H$) and the zero-field specific heat data using the relation:~\cite{deltaT1,deltaT2}
\begin{eqnarray}
\Delta T_{ad} (T,H) = - \frac{T \Delta S_{m}}{C_{p}(T,H)}.
\end{eqnarray}
The peak value of $\Delta T_{ad}$ is 7.5 K for a field change of 9 T, as shown in Fig.~\ref{Fig6}(d).

For a refrigeration cycle, the amount of heat transferred from hot to cold sinks can be estimated by the cooling power per unit volume, simply known as RC and RCP. The RC and RCP can be evaluated from $- \Delta S_m$ vs $T$ curve using the mathematical formula $RC~=~\int_{T_1}^{T_2} |\Delta S_m| dT$ and RCP~$=$~$|- \Delta S^{max}_m|$~$\times$~$\delta T_{FMHM}$, respectively ~\cite{rev3,RC,Eu4PdMg}. Here, $\rm{T_{1}}$ and $\rm{T_{2}}$ are the temperatures corresponding to the left and right sides of the half maximum $- \Delta S_m$ peak. $\delta T_{FMHM}$ is the full width at half maximum for each field. The obtained values of RC (RCP) are found to be as large as 103(140) J/kg and 302(415) J/kg for the changes of magnetic field 0$\textendash$5 T and 0$\textendash$9 T, respectively. The obtained values of $\mathrm{Gd_2AgSi_3}$ are compared in Table~\ref{compare}. As seen from Table~\ref{compare}, the obtained MCE parameters of $\mathrm{Gd_{2}AgSi_{3}}$ are quite small in comparison with other reported MCE materials.

\subsection{\label{sec:level2}Summary}

In summary, we have experimentally studied the crystal structure, magnetic, transport, magnetoresistance and magnetic cooling properties of a new stoichiometric compound $\rm{Gd_{2}AgSi_{3}}$. This polycrystalline sample successfully formed as a single phase in a tetragonal $\alpha$-$\rm{ThSi_{2}}$-type crystal structure with space group ${I_{{4}_{1}}/amd}$. 
Magnetic susceptibility, electrical resistivity and specific heat measurements reveal that the compound exhibits two antiferromagnetic transitions at 11 K and 20 K. The superzone effect from resistivity is observed at below the transition temperature.
From field dependent isotherm resistivity characterization, it is confirmed that the compound possess negative magnetoresistance in the both antiferromagnetic ordered and paramagnetic region. 
The magnetic field induced first-order magnetic transitions in the magnetically ordered state is confirmed from the Arrott-plots. MCE of this present compound is systematically studied and the obtained values are compared with other reported compounds. These results contribute towards a better understanding of this class of materials.

\subsection*{\label{sec:level3}Acknowledgements}   
This work is supported by Global Excellence and Stature (UJ-GES) fellowship, University of Johannesburg, South Africa. DFR thanks OWSD and SIDA for the fellowship towards PhD studies. AMS thanks the URC/FRC (935A9) of UJ for assistance of financial support.

\subsection*{\label{ref}References}

\end{document}